\pdfoutput=1

\documentclass[10pt]{article}
\usepackage{graphicx}
\usepackage{amsmath}
\usepackage[bbgreekl]{mathbbol}
\usepackage{bbm}
\usepackage{amsfonts}
\usepackage{amssymb}

\begin{document}
\title{Note on invariant properties of a quantum system placed into thermodynamic environment  \\ {\small (Published: Physica A 398 (2014) 65–-75) } }
\author{A. Y. Klimenko\thanks{Email for communications: klimenko@mech.uq.edu.au  } \\ 
{\it The University of Queensland, SoMME} \\ {\it QLD 4072, Australia} \\  \bigskip
}



\maketitle

\begin{abstract}
The analysis conducted in this work indicates that interactions of a CP-violating (and CPT-preserving) quantum system with a thermodynamic environment 
can produce the impression of a CPT violation in the system. 
This conclusion is reasonably consistent with the results reported for decays of neutral K-mesons.
\end{abstract}


%
%
%
%
%
%
%
%
%
%
%
%
%
%
%
%
%
%
%
%
%
%
%
%
%
%
%
%
%
%
%
%
%
%
%
%
%
%
%
%
%
%
%
%
%
%
%
%
%
%
%
%
%
%
%
%
%
%
%
%
%
%
%
%
%
%
%
%
%
%
%
%
%
%
%
%
%
%
%

\section{Introduction}

The influence of a thermodynamic environment, commonly referred to as a
thermodynamic bath, on quantum systems has been repeatedly discussed in
publications. Zurek \cite{Zurek1982} introduced a theory explaining loss of
coherence in a quantum system under the influence of the environments that
have a large number of degrees of freedom. Goldstein et. al. \cite{CT-G2006},
Popescu et. al. \cite{CT-P2006} and others demonstrated the property called
canonical typicality: for most pure states of the environment, the quantum
system behaves as if the environment was in the thermodynamic (i.e. maximally
mixed) state. Linden et. al. \cite{CT-P2009} proved that under certain
conditions the evolution of a quantum system placed into a bath leads to
equilibration. These and other aspects of thermalisation have been reviewed by
Yukalov \cite{Yukalov2012}, who stressed that, practically, any quantum system
cannot be completely isolated and is subject to some influence from the
environment. The existence of a degree of similarity between thermodynamic and
pure-state quantum engines has been discussed by Abe \cite{Abe2011}.

The goal of the present work is to show that thermodynamic environment can
affect apparent invariant properties of a quantum system, whose intrinsic
behaviour involves a CP-violation. Here we refer to charge, parity and time
symmetries conventionally denoted by C, P and T in quantum mechanics
\cite{Symmetry1972}. While the overwhelming majority of known quantum effects
are CP-compliant, the case of CP violation in decay of K-mesons (kaons) has
been known and investigated for many decades \cite{Symmetry1976}.

While the present work focuses on the effects induced by the environment
(assuming conventional unitarity of quantum evolutions) the likelihood of
spontaneous (intrinsic) violations of quantum mechanics, which can coexist
with induced mechanisms and also be responsible for thermalisation, has been
repeatedly discussed in publications \cite{Beretta2005, Zurek2002, QTreview}.
The possibility that these violations can affect decays of K-mesons has been
raised by Ellis et. al. \cite{QMviol1984}. Although the influence of
spontaneous violations of quantum mechanics can not be excluded a priori and
needs to be considered, such violations remain outside the scope of the
present work. Here, we consider the influence of the environment within the
framework of conventional quantum mechanics while relying on causality and
proper choice of parameters to reproduce the thermodynamic direction of time.

\section{Quantum system in a radiation bath}

Consider a quantum system, which involves both particles and antiparticles and
is placed into environment filled by radiation. The radiation is equilibrated
by surroundings, which, of course, are made of matter prevalently present in
the Universe. The system under consideration is a quantum system but is
subject at least to some thermodynamic influence from the environment. Since
antiparticles cannot interact weakly (i.e. without annihilations) with an
environment formed by matter, these interactions are performed only through
radiation, which is always generated by surrounding matter having non-zero
temperatures. Note that we do not consider stronger interaction of quantum
system with radiation through emission or adsorption. Only very weak
interactions of the system and radiation that tend to impose quantum
decoherence on the system \cite{Zurek1982,Yukalov2012} are of interest, while
radiation refers to any field that can be responsible for such interactions.
The system placed in a radiation bath, which is equilibrated by surroundings,
is schematically depicted in Figure 1. This scheme of interactions of the
system and the environment through the radiation bath reflects the fact that
surrounding matter needs to be removed from direct contact with the system
that involves both particles and antiparticles. We conduct our analysis within
the limits of quantum mechanics and take into account the thermodynamic
direction of time through causality and choice of the interaction parameters.

\begin{figure}[h]
\begin{center}
\includegraphics[width=8.5cm,trim=8cm 4cm 8cm 2cm,clip]{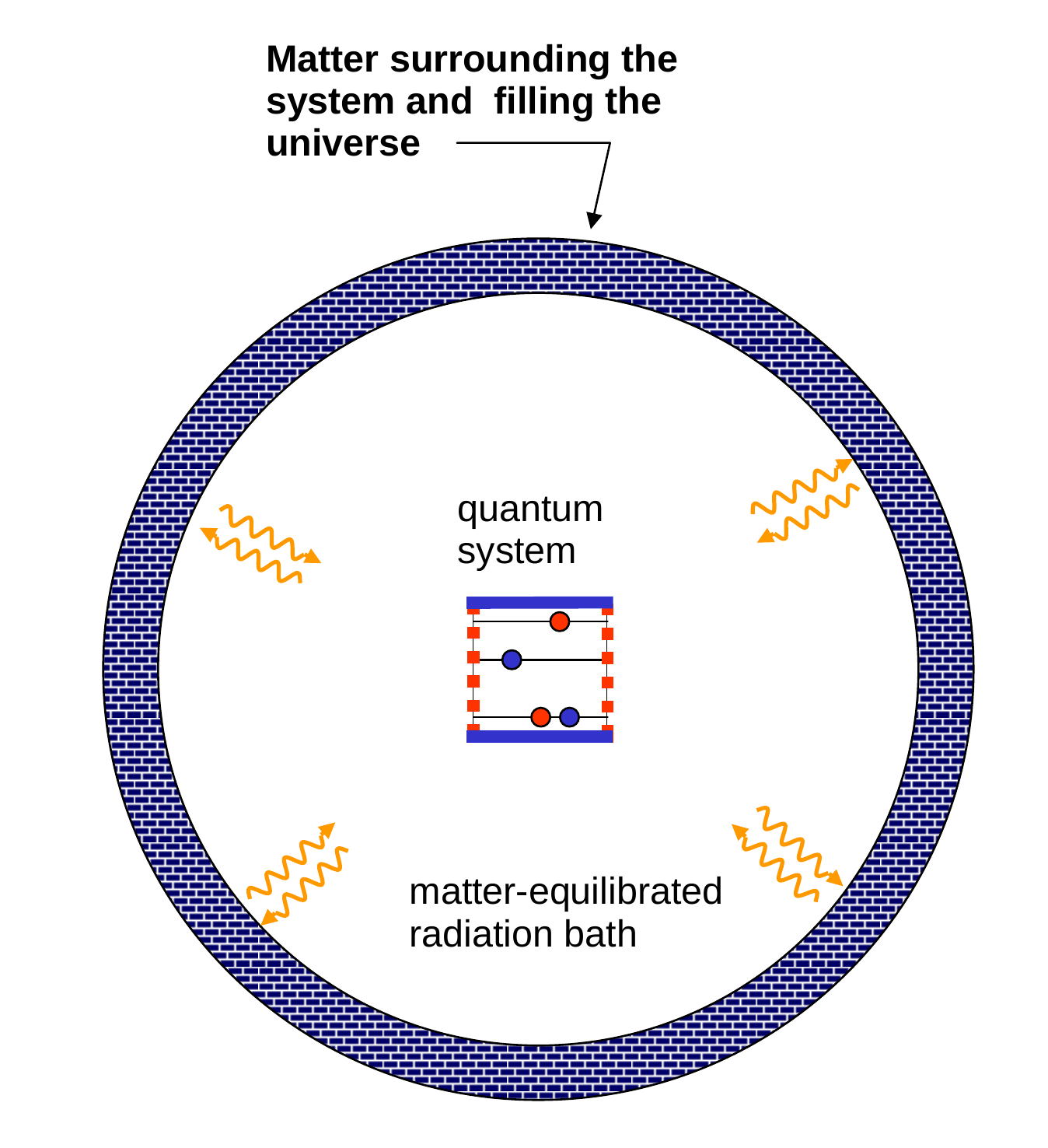}
\caption{Quantum system placed into a radiation bath.}
\label{fig6}
\end{center}
\end{figure}

The state of the radiation bath is characterised by its set of energy
eigenstates $\mathbb{H}_{B}\left|  \beta\right\rangle =E_{\beta}\left|
\beta\right\rangle .$ The dimension of this system is very large. In the same
way, the state of the environment, which, as some publications \cite{CT-P2006}
prefer to describe, may involve the rest of the universe, is characterised by
an even larger set of eigenstates $\mathbb{H}_{\Omega}\left|  \Omega
\right\rangle =E_{\Omega}\left|  \Omega\right\rangle .$ The state of the
system can be described by a set of orthogonal ket states $\left|
s\right\rangle $. Hence the overall state of the universal system involving
the system, the bath and the environment is specified by vectors in the tensor
product space $\left|  s\right\rangle \otimes\left|  \beta\right\rangle
\otimes\left|  \Omega\right\rangle $ (and the corresponding bra space
$\left\langle s\right|  \otimes\left\langle \beta\right|  \otimes\left\langle
\Omega\right|  $). Our analysis is conducted under several assumptions, which
can be summarised by:

\begin{enumerate}
\item  the quantum system is small and connected to a much larger environment
through the radiation bath;

\item  the radiation bath can exercise some influence on the system;

\item  the system has little effect on the bath, which is equilibrated by the
environment and remains in a thermodynamic state.
\end{enumerate}

Our use of the words ''strong'' and ''weak'' generally pertains to common
understanding of these terms, which nevertheless does not exclude links with
the strong and electroweak interactions of particle physics. Although the
presence of the environment must always be kept in mind, our assumptions lead
to autonomous consideration of the supersystem, which involves only the system
and the radiation bath. The state of the supersystem is specified by the
tensor product, which can be equivalently denoted by $\left|  s\right\rangle
\otimes\left|  \beta\right\rangle =\left|  s\right\rangle \left|
\beta\right\rangle =\left|  s\beta\right\rangle $. Note that we do not intend
to demonstrate that the bath should be in its thermodynamic state but simply
introduce this physical fact as a principal postulate of our analysis.

\subsection{The interaction Hamiltonian}

The overall Hamiltonian of the supersystem can be conventionally written in
the form%
\begin{equation}
\mathbb{H}=\mathbb{H}_{S}\otimes\mathbb{I}_{B}\mathbb{+I}_{S}\otimes
\mathbb{H}_{B}+\mathbb{H}_{SB} \label{H1}%
\end{equation}
where $\mathbb{H}_{S}$ is Hamiltonian of the system, $\mathbb{H}_{B}$ is the
Hamiltonian of the bath, $\mathbb{H}_{SB}$ is the system/bath interaction
Hamiltonian and $\mathbb{I}$ represents the corresponding identity operators.
Our assumptions, as detailed below, correspond to the following form of the
bath and interaction Hamiltonians
\begin{equation}
\left\langle \alpha\right|  \mathbb{H}_{B}\left|  \beta\right\rangle
=E_{\beta}I_{\alpha\beta},\;\;\;\left\langle q\alpha\right|  \mathbb{H}%
_{SB}\left|  s\beta\right\rangle =h_{qs}^{(\beta)}I_{\alpha\beta} \label{HSB}%
\end{equation}
Here, $I_{\alpha\beta}$ is the identity matrix $I_{\alpha\beta}=0$ for
$\alpha\neq\beta$ and $I_{\alpha\beta}=1$ for $\alpha=\beta$. The
corresponding indices run over the same sets of states, that is $\{s\}=$
$\{q\}$ and $\{\beta\}=$ $\{\alpha\}$.

The Hamiltonian specified by (\ref{HSB}) does not change the state of the bath
but may alter the behaviour of the system. Indeed, the supersystem wave
function $\Psi$ can be written in the form of the sum
\begin{equation}
\left|  \Psi(t)\right\}  =\sum_{\beta}p_{\beta}^{1/2}\left[  \beta\right]
\left|  \Psi_{\beta}(t)\right\rangle \label{PSI}%
\end{equation}
where
\begin{equation}
\left|  \Psi_{\beta}(t)\right\rangle =\left|  \psi^{(\beta)}\right\rangle
\left|  \beta\right\rangle ,\;\;\left|  \psi^{(\beta)}\right\rangle =\sum
_{s}c_{s\beta}(t)\left|  s\right\rangle \label{PSIbet}%
\end{equation}
Dependence on time can be omitted for brevity when this does not cause
confusion, for example $\left|  \Psi\right\}  =\left|  \Psi(t)\right\}  $.
Here, the random phases $\left[  \beta\right]  $ indicate that $\left|
\Psi\right\}  $ is not a pure state but is a mixture of pure states $\left|
\Psi_{\beta}\right\rangle $ with probabilities $p_{\beta}$. The notations
involving the random phases and their links with the density matrix
(operator)
\begin{equation}
\bbrho(t)=\left|  \Psi\right\}  \left\{  \Psi\right|  =\sum_{s,q,\beta
}p_{\beta}\;c_{s\beta}(t)c_{q\beta}^{\ast}(t)\left|  s\beta\right\rangle
\left\langle q\beta\right|  \label{rhoo}%
\end{equation}
are explained in Appendix A. The shape of the density matrix $\bbrho$ depends
on presence of $\left[  \beta\right]  $ in (\ref{PSI}), which forces
$\beta=\alpha$ by $\left\langle \left[  \alpha\right]  ^{\ast}\left[
\beta\right]  \right\rangle =I_{\alpha\beta}$ in (\ref{rhoo}).

Hamiltonian (\ref{HSB}) corresponds to our assumption 3 as well as to the
action of quantum decoherence \cite{Zurek1982,Yukalov2012} and canonical
typicality \cite{CT-G2006,CT-P2006}. This Hamiltonian ensures that not only
$\left|  \Psi\right\}  $ but also every $\left|  \Psi_{\beta}\right\rangle $
represents a solution of the Schrodinger equation for the supersystem
\begin{equation}
i\frac{\partial}{\partial t}\left|  \Psi_{\beta}(t)\right\rangle
=\mathbb{H}\left|  \Psi_{\beta}(t)\right\rangle \label{PSIbH}%
\end{equation}
For simplicity, the time $t$ is measured in Planck constants $\hbar$. If
$\left|  \Psi_{\beta}\right\rangle $ was not an autonomous solution, each
$\left|  \Psi_{\beta}\right\rangle $ would evolve into a superposition of pure
states involving different $\left|  \beta_{1}\right\rangle ,$ $\left|
\beta_{2}\right\rangle ,...$ This would contradict our understanding
\cite{Zurek1982,CT-G2006,CT-P2006,CT-P2009,Yukalov2012} that interactions of
the bath with the environment should decohere the bath eigenstates $\left|
\beta_{1}\right\rangle ,$ $\left|  \beta_{2}\right\rangle ,...$ and dissolve
any superposition of these states into a mixture. The presence of a small
system should not change the behaviour of the bath. The wave function of the
bath is deemed to stay in thermal equilibrium specified by the mixed state
\begin{equation}
\left|  \Psi_{B}\right\}  =\sum_{\beta}p_{\beta}^{1/2}e^{-iE_{\beta}t}\left[
\beta\right]  \left|  \beta\right\rangle ,\;\;\;p_{\beta}\sim\exp\left(
\frac{-E_{\beta}}{k_{B}T}\right)  \label{PSIB}%
\end{equation}
Note that although the eigenstates of the bath are not affected by
interactions with the system according to (\ref{HSB}), the eigenvalues
$E_{\beta}$ can in principle be altered by these interactions. This
alteration, however, is practically insignificant and neglected here ---
consider that the bath is more affected by the environment than by the system.

\subsection{Reduced density matrix and wave function of the system}

The behaviour of the quantum system is conventionally characterised by the
reduced density operator
\begin{equation}
\bbrho_{S}(t)=\sum_{\beta}\left\langle \beta\right|  \bbrho\left|
\beta\right\rangle =\sum_{s,q}\left(  \rho_{S}\right)  _{sq}\left|
s\right\rangle \left\langle q\right|  ,\;\;\left(  \rho_{S}\right)  _{sq}%
=\sum_{\beta}p_{\beta}c_{s\beta}(t)c_{q\beta}^{\ast}(t), \label{rho}%
\end{equation}
which is obtained from the density $\bbrho$ specified in (\ref{rhoo}) by
tracing the bath states out. One can easily see that the mixed state of the
system specified by the wave function
\begin{equation}
\left|  \Psi_{S}(t)\right\}  =\sum_{\beta}p_{\beta}^{1/2}\left[  \beta\right]
\left|  \psi^{(\beta)}\right\rangle , \label{PSI_S}%
\end{equation}
where $\left|  \psi^{(\beta)}\right\rangle ,$ given by (\ref{PSIbet}), is
equivalent to the density operator specified by (\ref{rho}). That is
$\bbrho_{S}=\left|  \Psi_{S}\right\}  \left\{  \Psi_{S}\right|  ,$ where the
random phase rule $\left\langle \left[  \alpha\right]  ^{\ast}\left[
\beta\right]  \right\rangle =I_{\alpha\beta}$ applies. This equivalence
implies that, for any system observable $Q_{S},$ the values found from the
density matrix and from the wave function are indistinguishable to the
observer%
\begin{equation}
Q_{S}=\operatorname{tr}\left(  \bbrho_{S}\mathbb{Q}_{S}\right)  =\left\{
\Psi_{S}\right|  \mathbb{Q}_{S}\left|  \Psi_{S}\right\}  =\sum_{\beta}%
p_{\beta}Q_{S}^{(\beta)},\;\;\;Q_{S}^{(\beta)}=\left\langle \psi^{(\beta
)}\right|  \mathbb{Q}_{S}\left|  \psi^{(\beta)}\right\rangle
\end{equation}
and so are the mixed states specified by (\ref{rho}) and (\ref{PSI_S}). Hence,
we can use $\left|  \Psi_{S}(t)\right\}  $, which is called here the reduced
wave function, as a convenient tool for investigating of the state of the
system. Note that in this case the reduced wave function $\left|  \Psi
_{S}(t)\right\}  $ specified by (\ref{PSI_S}) can be obtained from the wave
function of the supersystem $\left|  \Psi(t)\right\}  $ specified by
(\ref{PSI}) through the following procedure: multiplying $\left|
\Psi\right\}  $ by the bra $\left\langle \alpha\right|  ,$ summing over all
$\alpha$
\begin{equation}
\left|  \Psi_{S}(t)\right\}  =\sum_{\alpha}\left\langle \alpha||\Psi\right\}
=\sum_{\alpha,\beta}\left\langle \alpha|\beta\right\rangle p_{\beta}%
^{1/2}\left[  \beta\right]  \left|  \psi^{(\beta)}\right\rangle
\end{equation}
and noting orthonormality $\left\langle \alpha|\beta\right\rangle
=I_{\alpha\beta}$. The double ''$||$'' indicates incompleteness of the inner
product since the ket spans over a wider space $\left|  \Psi\right\}
\sim\left|  s\right\rangle \left|  \beta\right\rangle $ than the bra
$\left\langle \alpha\right|  $.

The system wave functions $\left|  \Psi_{S}^{(\beta)}\right\rangle ,$ which
are similar to $\left|  \psi^{(\beta)}(t)\right\rangle $ but have the effect
of the bath energy eigenstates $E_{\beta}$ factored out, are introduced by
\begin{equation}
\left|  \psi^{(\beta)}(t)\right\rangle =e^{-iE_{\beta}t}\left|  \Psi
_{S}^{(\beta)}(t)\right\rangle , \label{PSI_be}%
\end{equation}
that is%
\[
\left|  \Psi_{S}^{(\beta)}(t)\right\rangle =\sum_{s}C_{s}^{(\beta)}(t)\left|
s\right\rangle ,\;\;\;C_{s}^{(\beta)}=e^{iE_{\beta}t}c_{s\beta}%
\]
The wave functions $\left|  \Psi_{S}^{(\beta)}\right\rangle $ can now be
considered independently from each other. Note that the expression for the
reduced wave function remains the same
\begin{equation}
\left|  \Psi_{S}(t)\right\}  =\sum_{\beta}p_{\beta}^{1/2}\left[  \beta^{\circ
}\right]  \left|  \Psi_{S}^{(\beta)}(t)\right\rangle ,\;\;\;\left[
\beta^{\circ}\right]  =e^{-iE_{\beta}t}\left[  \beta\right]  \label{S_PSIb}%
\end{equation}
since, as discussed in the Appendix, the random phases $\left[  \beta^{\circ
}\right]  $ are functionally equivalent to $\left[  \beta\right]  .$\ Each
$\left|  \Psi_{S}^{(\beta)}\right\rangle $ represents an independent solution
of the reduced Schrodinger equation
\begin{equation}
i\frac{\partial}{\partial t}\left|  \Psi_{S}^{(\beta)}(t)\right\rangle
=\mathbb{H}^{(\beta)}\left|  \Psi_{S}^{(\beta)}(t)\right\rangle \label{S_PSIH}%
\end{equation}
with the $\beta$-dependent effective system Hamiltonian $\mathbb{H}^{(\beta)}$
specified by
\begin{equation}
\mathbb{H}^{(\beta)}=\left\langle \beta\right|  \mathbb{H}\left|
\beta\right\rangle -E_{\beta}\mathbb{I}_{S}=\mathbb{H}_{S}+\mathbbm{h}%
^{(\beta)} \label{S_H}%
\end{equation}%
\begin{equation}
\mathbbm{h}^{(\beta)}=\left\langle \beta\right|  \mathbb{H}_{SB}\left|
\beta\right\rangle =\sum_{s,q}h_{sq}^{(\beta)}\left|  s\right\rangle
\left\langle q\right|  \label{H3}%
\end{equation}
Equation (\ref{S_PSIH}) can be obtained by multiplying (\ref{PSIbH}) by
$\left\langle \beta\right|  \left[  \beta\right]  ^{\ast}$ and factorising
$e^{i(\mathbb{I}_{S}\otimes\mathbb{H}_{B})t}$ from the rest of the
solution.\ Here, we take into account that the operators $\mathbb{H}$ and
$\mathbb{I}_{S}\otimes\mathbb{H}_{B}$ commute $\left[  \mathbb{H}%
,\mathbb{I}_{S}\otimes\mathbb{H}_{B}\right]  =0$ as long as $\mathbb{H}_{B}$
and $\mathbb{H}_{SB}$ are defined by (\ref{HSB}). Note that the operators
$\mathbb{H}$ and $\mathbb{H}_{S}\otimes\mathbb{I}_{B}$ are not necessarily
commutative due to the possibility of $\left[  \mathbb{H}_{S},\mathbbm
{h}^{(\beta)}\right]  \neq0$.

\subsection{On unitarity of quantum models and decoherence}

From the perspective of the canonical typicality, the state of the bath ---
maximally mixed or a typical pure state --- should have little effect on the
properties of the system characterised by reduced quantities. While being
important, this property does not change our treatment since the bath is
physically in thermodynamic equilibrium with the environment. When considered
from the perspective of interactions of the small system with the large bath,
the microcanonical maximally mixed state of the bath (i.e. state with maximal
entropy and fixed total energy) is essentially the same as the canonical
equilibrium state of the bath specified by (\ref{PSIB}). The main effect of
the bath on the state of the system is in inducing decoherence of the
eigenstates of the system, which is not the same as decoherence of the wave
functions multiplied by different random phases $\left[  \beta\right]  $ in
(\ref{PSI}) and (\ref{PSI_S}). Assuming that the states $\left|
s\right\rangle $ are energy eigenstates of the system in the absence of
interference from the bath (this is assumed only in this subsection), the
decoherence of these eigenstates results in
\begin{equation}
\left(  \rho_{S}\right)  _{sq}\approx0\text{ for }s\neq q
\end{equation}
that is the reduced density matrix becomes diagonal. This property is not
synonymous with the decoherence of $\left[  \beta\right]  \left|  \Psi
_{S}^{(\beta)}\right\rangle $.

The theory of quantum decoherence induced by interactions with a bath or
environment has been introduced by Zurek \cite{Zurek1982}. This theory is
based on assuming that the interaction term is diagonal, that is
\begin{equation}
h_{sq}^{(\beta)}=h_{s}^{(\beta)}I_{sq} \label{h_d}%
\end{equation}
implying that operators $\mathbb{H}_{S}$ and $\mathbbm{h}^{(\beta)}$ commute
\begin{equation}
\left[  \mathbb{H}_{S},\mathbbm{h}^{(\beta)}\right]  =0 \label{com}%
\end{equation}
Note that (\ref{h_d}) and (\ref{com}) are assumptions of the conventional
decoherence theory \cite{Zurek1982} and not of the present work. The solution
of equation (\ref{S_PSIH}) is then easily obtainable
\begin{equation}
\left|  \Psi_{S}^{(\beta)}(t)\right\rangle =\exp\left(  -i\left(  E_{s}%
+h_{s}^{(\beta)}\right)  t\right)
\end{equation}
resulting in
\begin{equation}
\left(  \rho_{S}\right)  _{sq}=A_{sq}\exp\left(  -i\left(  E_{s}-E_{q}\right)
t\right)  ,\;\;\;A_{sq}=\sum_{\beta}p_{\beta}\exp\left(  -i\left(
h_{s}^{(\beta)}-h_{q}^{(\beta)}\right)  t\right)  \label{Asq}%
\end{equation}
As a sum of a large number of uncorrelated values, $A_{sq}$ should be small on
average assuming that the number of probable states indexed by $\beta$ is
large and, of course, that $q\neq s$ \cite{Zurek1982}.

While decoherence is an essentially non-unitary (and irreversible) process
increasing the entropy, the description of induced decoherence given above is
based on placing the system into a bath to form a supersystem of a much larger
dimension than that of the system. However, even a very large supersystem is
still described by its Schrodinger equation (\ref{PSIbH}), which is unitary
and, consequently, time reversible. Is this reversible description consistent
with the irreversible reality? This question can be traced back to historic
discussions between Boltzmann, Loschmidt and Zermelo. It should be noted first
that both canonical typicality and induced decoherences deal with average,
typical properties that may, without contradicting the formulation of these
statements, be subject to some rare exceptions. For example, $A_{sq}(t)$ is a
quasiperiodic function of time in (\ref{Asq}) and, eventually, should recur to
the proximity of its initial value $A_{sq}(0)$, but, in large supersystems,
these recurrences occur after extremely long times and would be impossible to
observe in practice. A very small correction, uncertainty or external
interference added to the model would be sufficient to remove these
recurrences. While, as remarked in the Introduction, spontaneous violations of
unitarity are also likely to play a role in decoherence and thermalisation,
the present work focuses on the induced mechanisms of decoherence. Unitary
evolution of a sufficiently large dimension combined with the causality
principle can provide a very good approximation for irreversible evolutions
observed in reality.

\subsection{Decay kinetics}

This section introduces an influence of the environment into the
Weisskopf-Wigner approximation \cite{WWA1930}. Since the 1930s, this
approximation is conventionally used for characterisation of decaying states.
The Weisskopf-Wigner approximation is derived from the Schrodinger equation
for specific initial conditions and under assumptions of relatively weak
interactions of the few initial with many final states. Later publications
\cite{WWA1974} demonstrated the rigorous asymptotic character of this approximation.

Among the system states $\left|  s\right\rangle $, we distinguish two groups:
the initial state or states $\left|  k\right\rangle $ ($\left\langle j\right|
$ is used when an alternative index is needed for the bra-space), and many
possible final states, which are indexed by $\left|  f\right\rangle ,$ that is
$\{s\}=\{k,f\}$ and $\{k\}=\{j\}$. The system is initially placed into a pure
state that spans over one or several $k$-states but then decays into one or
several of the $f$-states. The system Hamiltonian is given by\ $\mathbb{H}%
_{S}=\mathbb{H}_{0}+\mathbb{H}_{1}$, where the smaller component
$\mathbb{H}_{1}$ is responsible for relatively weak interactions of the
initial and final states. The states $\left|  s\right\rangle ,$ involving the
initial $\left|  k\right\rangle $ and final $\left|  f\right\rangle $ states,
are eigenstates $\mathbb{H}_{0}\left|  s\right\rangle =E_{s}\left|
s\right\rangle $ of the undisturbed Hamiltonian $\mathbb{H}_{0}$ (i.e. strong
eigenstates) but not necessarily the eigenstates of $\mathbb{H}_{S}$. If there
are several initial states (for example a particle and its antiparticle), then
these states are assumed to possess the same energy $E_{k}=E_{0}$ for all $k$.
After tracing out the state of the environment from the Hamiltonian, we
rewrite (\ref{H1}) as
\begin{equation}
\mathbb{H}^{(\beta)}=\left\langle \beta\right|  \mathbb{H}\left|
\beta\right\rangle -E_{\beta}\mathbb{I}_{S}=\mathbb{H}_{0}+\mathbb{H}%
_{2}^{(\beta)},\;\;\mathbb{H}_{2}^{(\beta)}\equiv\mathbb{H}_{1}+\mathbbm
{h}^{(\beta)} \label{H2}%
\end{equation}

The matrix $\Lambda_{kj}^{(\beta)},$ which specifies the evolution of the
$k$-states
\begin{equation}
i\frac{d}{dt}\left|  \Psi_{k}^{(\beta)}\right\rangle =\sum_{j}\Lambda
_{kj}^{(\beta)}\left|  \Psi_{j}^{(\beta)}\right\rangle
\end{equation}
is obtained by applying the Weisskopf-Wigner approximation \cite{WWA1930} to
equation (\ref{S_PSIH})%
\begin{equation}
\Lambda_{jk}^{(\beta)}=\left\langle j\right|  \mathbb{H}^{(\beta)}\left|
k\right\rangle +\lambda_{jk}^{(\beta)},\;\;\;\lambda_{jk}^{(\beta)}\equiv
\sum_{f}\frac{\left\langle j\right|  \mathbb{H}_{2}^{(\beta)}\left|
f\right\rangle \left\langle f\right|  \mathbb{H}_{2}^{(\beta)}\left|
k\right\rangle }{E_{0}-E_{f}+i\varepsilon} \label{WW}%
\end{equation}
where $\varepsilon\rightarrow0$\ and the sign of $\varepsilon$ is selected to
produce decaying exponents as required by causality. The matrix $\Lambda
_{jk}^{(\beta)}$ is generally not Hermitian, while Hermitian $M_{jk}^{(\beta
)}$ and $\Gamma_{jk}^{(\beta)}$ in
\begin{equation}
\Lambda_{jk}^{(\beta)}=M_{jk}^{(\beta)}-\frac{i}{2}\Gamma_{jk}^{(\beta)}
\label{LMG}%
\end{equation}
represent the energy-mass matrix and the decay matrix correspondingly. The
imaginary component of the sum $\lambda_{jk}^{(\beta)},$ which is associated
with the term $\Gamma_{jk}^{(\beta)},$ appears as the contribution of
so-called ''on-shelf'' states $\hat{f}$\ where $E_{0}=E_{\hat{f}}.$ The states
$\check{f},$ representing the remaining or ''off-shelf'' states in
$\{f\}=\{\hat{f},\check{f}\}$, contribute only to $M_{jk}^{(\beta)}$. Since
the set of final states is large and, possibly, continuous, the sum in
(\ref{WW}) is understood and evaluated as the corresponding integral.

\subsection{The interaction term and the eigenstates}

According to Zurek's theory \cite{Zurek1982}, the terms responsible for
decoherence act primarily along the eigenstates of the system --- see the
discussion around equation (\ref{com}). Applying the same strategy to the
present case encounters difficulties for the on-shelf states that are
discussed below. These difficulties indicate that (\ref{com}) is excessively
restrictive and not suitable for the present analysis.

If we restrict our attention only to the strongest interactions specified by
the undisturbed Hamiltonian $\mathbb{H}_{0}$ then at the leading order$,$ this
implies that $\mathbbm{h}^{(\beta)}$ is diagonal
\begin{equation}
\left\langle q\right|  \mathbbm{h}^{(\beta)}\left|  s\right\rangle \approx
h_{s}^{(\beta)}I_{qs}%
\end{equation}
This expression is responsible for interference of the bath with the initial
$\left|  k\right\rangle $ states and final $\left|  f\right\rangle $ states
autonomously without any transition from $\left|  k\right\rangle $ to $\left|
f\right\rangle $, while we are interested in interference of the bath with the
process of the transition.

One might try to correct the set of eigenstates $\{\left|  s\right\rangle
\}=\{\left|  k\right\rangle ,\left|  f\right\rangle \}$ to account for weak
interactions specified by $\mathbb{H}_{1}$. At the leading order, the
corrected eigenstates $\left|  s^{\prime}\right\rangle $ are given by the
standard quantum perturbation theory \cite{LL} yielding
\begin{equation}
\left|  s^{\prime}\right\rangle =\left|  s\right\rangle +\sum_{q\neq s}%
\frac{\left\langle q\right|  \mathbb{H}_{1}\left|  s\right\rangle }%
{E_{s}-E_{q}}\left|  q\right\rangle +...
\end{equation}
Assuming that, according to (\ref{com}), the decohering term is aligned with
the corrected eigenstates $\left|  s^{\prime}\right\rangle $ we obtain
\begin{equation}
\mathbbm{h}^{(\beta)}=\sum_{s}h_{s}^{(\beta)}\left|  s^{\prime}\right\rangle
\left\langle s^{\prime}\right|  =\sum_{s}h_{s}^{(\beta)}\left|  s\right\rangle
\left\langle s\right|  +\sum_{s,q\neq s}\left(  h_{s}^{(\beta)}-h_{q}%
^{(\beta)}\right)  \frac{\left\langle q\right|  \mathbb{H}_{1}\left|
s\right\rangle }{E_{s}-E_{q}}\left|  q\right\rangle \left\langle s\right|
+... \label{hss}%
\end{equation}
Substitution of (\ref{hss}) and (\ref{H2}) into (\ref{WW}) produces integrals
(represented by sums over $f$ in the equations) which diverge as $(E_{0}%
-E_{f})^{-2}$ in vicinities of the on-shelf states $\hat{f},$ where
$E_{0}=E_{f}$. This shows that bath interference with decay is mostly
contributed to by the vicinities of the on-shelf states $\hat{f}$ and, at the
same time, indicates that our approximations and assumptions need to be
reevaluated in these vicinities. Since the exact sub-atomic mechanisms of
decoherence remain unknown, our consideration is necessarily based on
qualitative analysis of the system/bath interaction term.

Since the on-shelf states $\hat{f}$ are degenerate due to $E_{0}=E_{f}$, the
eigenstates experience large and rapid changes in vicinities of these states,
induced by higher order terms in the Hamiltonian \cite{LL} (the degeneracy of
the initial states $E_{k}=E_{0}$ is uniformly present in all transitions and
does not cause rapid adjustments). Although the interference terms are likely
to be affected when $f\rightarrow\hat{f}$, these terms can not be assumed to
be fully aligned with the rapidly changing eigenstates in vicinity of the
on-shelf states. Hence, the interference term is no longer compliant with
(\ref{com}), when the on-shelf states are involved. This implies interactions
between different energy modes. While the decoherence theory \cite{Zurek1982}
presumes decoherence at short times without energy exchange, decays are more
affected by interference with the bath and involves some redistribution of
energy in vicinity of the on-shelf states.

\section{Invariant properties}

We note that the overwhelming majority of quantum effects that we know of are
CP-invariant. Only very few exceptions have been found that contradict this
rule. We assume that the system under consideration may be one of these
exceptions, i.e. CP-violating and CPT-invariant. Note that here we refer only
to the intrinsic properties of the system and not to the interaction of the
system and the environment.

For interactions of the system and the thermodynamic environment, we assume CP
invariance and, possibly, T violation. These assumptions need further
comments. The current understanding of the influence of the environment on
quantum systems \cite{Zurek1982,CT-G2006,CT-P2006,CT-P2009,Yukalov2012} does
not have any provisions for discriminating between its effects on particles
and antiparticles; hence interactions of the system and the bath are expected
to be CP-invariant. As there is no proof or assertion that these interactions
must be T-invariant, we should allow for the possibility of T violations in
environmental interactions.

The last point needs a more detailed discussion. Thermodynamic interactions
taking place far from equilibrium are time directional and irreversible. For
example, decoherence always occurs only forward and not backward in time. A
number of theories \cite{Beretta2005, Zurek2002, QTreview,PenroseBook} relate
macroscopic irreversibility to intrinsic microscopic processes at quantum
level. These processes are expected to violate both quantum mechanics and
microscopic time symmetry while having very small magnitudes that make any
direct detection of the time-generating processes very difficult. In general,
these time-generating violations of unitarity should violate either CPT or CP
symmetry, giving rise to two possible alternative thermodynamics of
interactions of matter and antimatter \cite{KM_JETC2013}. In this work,
however, we do not consider violations of quantum mechanics and focus our
attention on the influence of the environment but note that the a priori
presumption of time symmetry in interactions of a quantum system and its
thermodynamic environment (which effectively encompasses the rest of the
universe) would not be justified.

One may notice that our interpretation of the system's interactions with the
universe --- CP-invariance combined with possible T-violation --- may
constitute a CPT-violation. In general, CPT invariance is a theorem in quantum
field theory but its validity in the context of statistical or general physics
would be only a hypothesis. While some published opinions (see, for example,
Penrose \cite{PenroseBook}) favour the view that the processes enacting the
second law of thermodynamics are CPT-violating and that CPT invariance is not
a property of the entire universe, we give a different interpretation for the
case considered here. If it exists, CPT violation in thermodynamic
interactions with the environment is only an apparent feature since the
operation of charge conjugation is not rigorous in our setup. Indeed, under
charge conjugation, we correctly swap particle and antiparticles in the system
and correctly leave the radiation bath unchanged but we do not and cannot
appropriately adjust the environment (i.e. change matter populating the
universe into antimatter). Thus our operator C is, strictly speaking,
incomplete, creating opportunities for apparent CPT violations when
interactions with the environment are of some influence. According to the line
of thinking adopted in quantum thermodynamics
\cite{Zurek1982,CT-G2006,CT-P2006,CT-P2009,Yukalov2012}, interactions with the
environment are deemed to be unavoidable.

Approximation (\ref{WW}) is conventionally used to analyse decay of neutral
kaons, $K^{\circ}$ and $\bar{K}^{\circ}$ \cite{Symmetry1976}. Here, we
investigate the CPT-compliance of the decay, which requires that the term
\begin{equation}
\Delta\Lambda^{(\beta)}=\Lambda_{KK}^{(\beta)}-\Lambda_{\bar{K}\bar{K}%
}^{(\beta)}%
\end{equation}
is nullified \cite{Symmetry1976}. Unlike the running indices $k$ and $j$, the
subscripts $K$ and $\bar{K}$ denote the fixed states corresponding to the kaon
and antikaon.

\subsection{Invariance of the system/bath interactions}

The property of radiation-induced interactions being CP-invariant and
T-asymmetric can be expressed in terms of the interaction Hamiltonian
\begin{gather}
\left\langle q\alpha\right|  \mathbb{H}_{SB}\left|  s\beta\right\rangle
=\left\langle \overline{q\alpha}\middle|\mathbb{H}_{SB}\middle|\overline
{s\beta}\right\rangle \Rightarrow h_{qs}^{(\beta)}=h_{\bar{q}\bar{s}}%
^{(\beta)}\label{CP}\\
\Delta h_{qs}^{(\beta)}=\left\langle q\beta\right|  \mathbb{H}_{SB}\left|
s\beta\right\rangle -\left\langle s\beta\right|  \mathbb{H}_{SB}\left|
q\beta\right\rangle \neq0 \label{TV}%
\end{gather}
where the overbars denote anti-states. The property $\beta=\bar{\beta}$ is
taken into account for radiation (i.e. matter and antimatter interact with
radiation in the same way). Note that $\Delta h_{qs}^{(\beta)}$ is imaginary
when interactions with the environment are unitary and $\mathbbm{h}^{(\beta)}$
is Hermitian (if $\mathbbm{h}^{(\beta)}$ were the term responsible for
violations of unitarity in thermodynamic interactions, this would correspond
to real $\Delta h_{qs}^{(\beta)}$). As previously noted, equations (\ref{CP})
and (\ref{TV}) imply an apparent CPT violation:
\begin{equation}
\left\langle q\beta\right|  \mathbb{H}_{SB}\left|  s\beta\right\rangle
\neq\left\langle \overline{s\beta}\right|  \mathbb{H}_{SB}\left|
\overline{q\beta}\right\rangle \label{CPTV}%
\end{equation}

\subsection{The case of a CP-invariant system}

Most quantum systems possess CP symmetry and relatively few CP violations are
known in quantum mechanics. When the system Hamiltonian $\mathbb{H}_{S}$ is
intrinsically CP-invariant, we obtain
\begin{align}
H_{kf}  &  =\left\langle k\right|  \mathbb{H}_{S}\left|  f\right\rangle
=\left\langle \overline{k}\right|  \mathbb{H}_{S}\left|  \overline
{f}\right\rangle =H_{\bar{k}\bar{f}}\label{CP2}\\
\Delta\Lambda^{(\beta)}  &  \equiv\Lambda_{KK}^{(\beta)}-\Lambda_{\bar{K}%
\bar{K}}^{(\beta)}=0 \label{CP3}%
\end{align}
Here we use equations (\ref{CP}),\ (\ref{WW}), (\ref{CP2}) to obtain
(\ref{CP3}) and conclude that the system appears to the observer as being
CPT-compliant. Here, we refer only to opinion of the observer, who uses
$\Delta\Lambda^{(\beta)}$ to test the CPT-invariance; in general, the system
may or may not be CPT-invariant under these conditions. We also note that,
since the Hamiltonians $\mathbb{H}_{0}$ and $\mathbb{H}_{2}^{(\beta)}$ are
CP-preserving, there is no T violation in this system apparent to the observer
$\Lambda_{K\bar{K}}^{(\beta)}=\Lambda_{\bar{K}K}^{(\beta)}$ , although, once
again, the system's interactions with the environment may in fact be T-violating.

\subsection{The case of a CP-violating and CPT-invariant system}

The system Hamiltonian $\mathbb{H}_{S}$ is CPT-invariant provided that%
\begin{equation}
H_{kf}=\left\langle k\right|  \mathbb{H}_{S}\left|  f\right\rangle
=\left\langle \overline{f}\right|  \mathbb{H}_{S}\left|  \overline
{k}\right\rangle =H_{\bar{f}\bar{k}} \label{CPT2}%
\end{equation}
We use equations (\ref{CP}),\ (\ref{WW}) and (\ref{CPT2}) and note that the
first term in (\ref{WW}) does not contribute to $\Delta\Lambda^{(\beta)}$ so
that
\begin{gather}
\Delta\Lambda^{(\beta)}=\lambda_{KK}^{(\beta)}-\lambda_{\bar{K}\bar{K}%
}^{(\beta)}%
\;\;\;\;\;\;\;\;\;\;\;\;\;\;\;\;\;\;\;\;\;\;\;\;\;\;\;\;\;\;\;\;\;\;\;\;\;\;\;\;\;\;\;\;\;\;\;\;\;\;\;\;\;\;\;\;\;\;\nonumber\\
=\sum_{f}\frac{\left(  H_{Kf}^{\prime}+h_{Kf}^{(\beta)}\right)  \left(
H_{fK}^{\prime}+h_{fK}^{(\beta)}\right)  -\left(  H_{\bar{K}\bar{f}}^{\prime
}+h_{\bar{K}\bar{f}}^{(\beta)}\right)  \left(  H_{\bar{f}\bar{K}}^{\prime
}+h_{\bar{f}\bar{K}}^{(\beta)}\right)  }{E_{0}-E_{f}+i\varepsilon}\nonumber\\
=\sum_{f}\frac{\left(  H_{Kf}^{\prime}+h_{Kf}^{(\beta)}\right)  \left(
H_{fK}^{\prime}+h_{fK}^{(\beta)}\right)  -\left(  H_{fK}^{\prime}%
+h_{Kf}^{(\beta)}\right)  \left(  H_{Kf}^{\prime}+h_{fK}^{(\beta)}\right)
}{E_{0}-E_{f}+i\varepsilon}\nonumber\\
=-\sum_{f}\frac{\Delta H_{Kf}^{\prime}}{E_{0}-E_{f}+i\varepsilon}\Delta
h_{Kf}^{(\beta)}%
\;\;\;\;\;\;\;\;\;\;\;\;\;\;\;\;\;\;\;\;\;\;\;\;\;\;\;\;\;\;\;\;\;\;\;\;\;
\label{CPT-v}%
\end{gather}
where we denote $H_{kf}^{\prime}\equiv\left\langle k\right|  \mathbb{H}%
_{1}\left|  f\right\rangle ,$ \ $\Delta H_{kf}^{\prime}=H_{kf}^{\prime}%
-H_{fk}^{\prime}$. As such, thermodynamic interactions with the environment
through radiation may appear in CP-violating (but CPT preserving) systems as
an apparent CPT violation. Absence of the time symmetry $\Delta H_{Kf}%
^{\prime}\neq0$ and $\Delta h_{Kf}^{(\beta)}\neq0$ is essential for this
effect. Note that the product $\Delta H_{Kf}^{\prime}\Delta h_{Kf}^{(\beta)}$
is always real, since $\Delta H_{Kf}^{\prime}$ and $\Delta h_{Kf}^{(\beta)}$
are both imaginary. As discussed in Section 2.4, the major contribution to the
sum (\ref{CPT-v}) takes place in the vicinities of the on-shelf states
$\hat{f}$ where $h_{sq}^{(\beta)}$ is not restricted to the diagonal $I_{sq}$
and $\Delta h_{sq}^{(\beta)}$ is non-zero.

\section{Tests of CPT invariance in kaon decays}

Decay of neutral kaons is one of very few known cases of CP-violation in
quantum systems (it was the only known case of CP-violation for several
decades and is widely covered in the literature see, for example,
\cite{Symmetry1976,PDG2012}). The CP violation takes the form $\Lambda
_{K\bar{K}}\neq\Lambda_{\bar{K}K},$ were we omit the superscript $\beta$ since
the CP violation is induced by $\Delta H_{kf}^{\prime}\neq0$ in (\ref{WW}) and
does not need any interaction with the environment (i.e. we can put
$h_{kf}^{(\beta)}=0$). This CP violation is conventionally understood as
preserving CPT and violating T invariance.

The CPT invariance of kaon decays has also been repeatedly tested
\cite{Symmetry1976,PDG2012}. The results of these tests are summarily
presented in a recent review of particle physics \cite{PDG2012} and are shown
in Figure 2. These results neither constitute an unambiguous CPT violation nor
assert the CPT invariance, although there is a noticeable CPT-violating bias
in the results. The magnitude of the apparent CPT violation is only 10 times
smaller than that of the established CP-violation in this system but the
relative scattering of the results is much larger in CPT than in CP tests
which prevents any conclusive statements on CPT invariance.

Our explanation of the bias is that the apparent CPT violation in kaon decays
is induced by thermodynamic interactions with the environment (and does not
contradict to a rigorous interpretation of CPT invariance). This violation can
be traced to the fact that, in our experiments, we cannot possibly perform
charge conjugation of the environment. As discussed below, this explanation
seems to be consistent with the results presented in Figure 2.

\begin{figure}[h]
\begin{center}
\includegraphics[width=10cm]{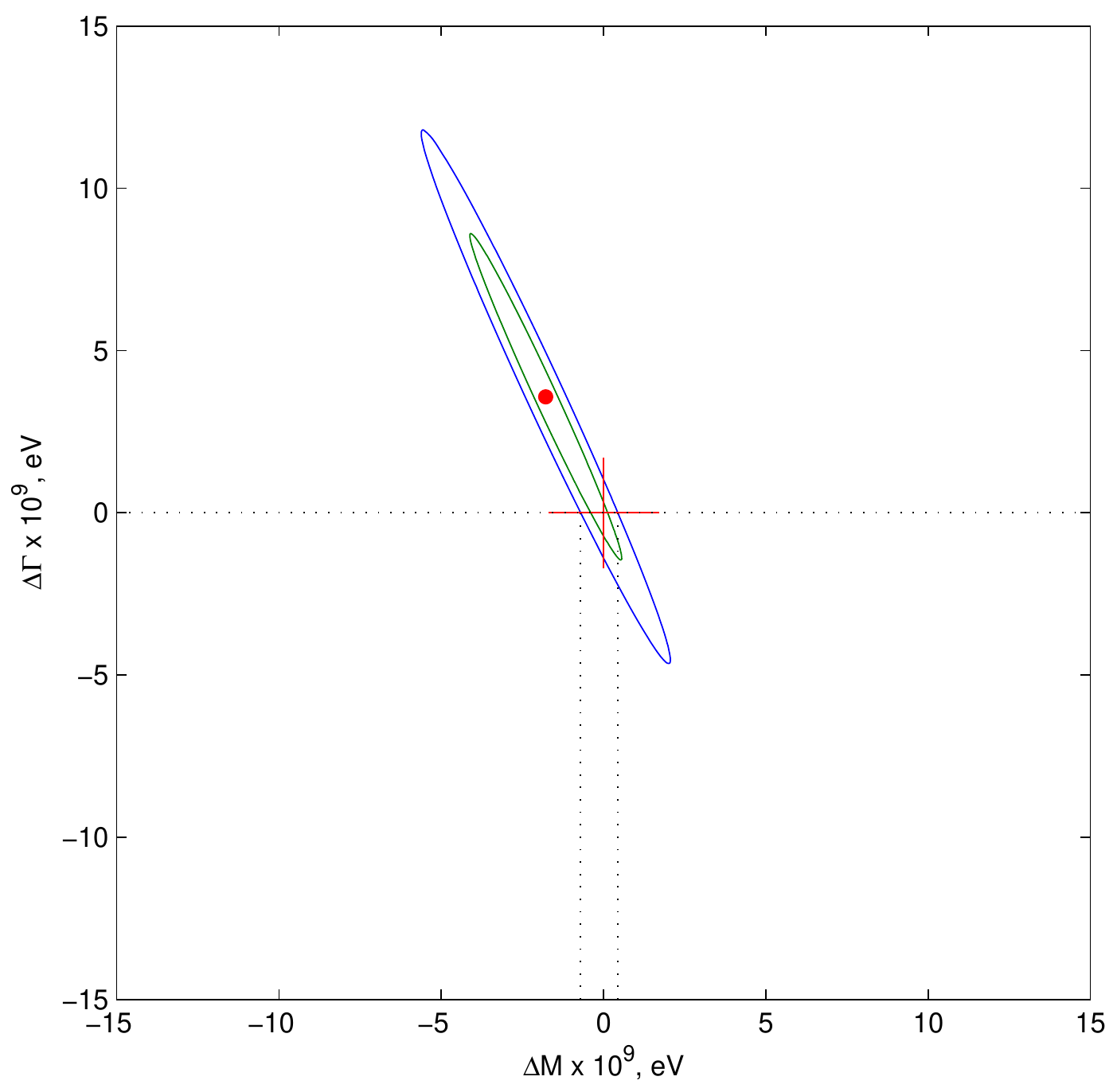}
\caption{Test of CPT invariance in kaon decays \cite{PDG2012}. The two contours show the 95\% and 65\% confidence levels. The solid dot corresponds to the mean value. The cross indicates the coordinate origin -- the point where no CPT violation occurs. The vertical dashed lines show the 95\% confidence region for $\Delta M$ in realisations conditioned on $\Delta\Gamma=0$.}
\label{figk}
\end{center}
\end{figure}

The apparent CPT violation in (\ref{CPT-v}) is caused by interference of the
environment with quantum interactions of the initial and the on-shelf final
states of the system. Hence, the contribution to the violation from the sum in
(\ref{CPT-v}) is dominated by the vicinities of the on-shelf states $\hat{f}$.
Since the product $\Delta H_{Kf}^{\prime}\Delta h_{Kf}^{(\beta)}$\ is always
real, this product contributes to both the real $\Delta M^{(\beta)}\equiv
M_{KK}^{(\beta)}-M_{\bar{K}\bar{K}}^{(\beta)}$ and imaginary $\Delta
\Gamma^{(\beta)}\equiv\Gamma_{KK}^{(\beta)}-\Gamma_{\bar{K}\bar{K}}^{(\beta)}$
components of $\Delta\Lambda^{(\beta)}$. This implies that, if only the
realisations $\beta$ that possess the property $\Delta\Gamma^{(\beta)}%
\approx0$ (due to $\Delta h_{Kf}^{(\beta)}$ being small in the vicinity of
$\hat{f}$ ) are selected, then the corresponding energy-mass term should also
vanish $\Delta M^{(\beta)}\approx0.$ This expectation is consistent with the
experimental results presented in Figure 2. Another feature of equation
(\ref{CPT-v}) is its linearity with respect to the fluctuating term $\Delta
h_{Kf}^{(\beta)},$ which does not contradict the results shown in Figure 2
that seem to indicate the existence of a linear correlation or connection
between $\Delta\Gamma^{(\beta)}$ and $\Delta M^{(\beta)}$. Fluctuations of
large magnitudes (i.e. of the same order as mean values) are also consistent
with the thermodynamic origin of the fluctuations, since microscopic systems
placed in a thermodynamic environment are subject to fluctuations of magnitude
comparable to the corresponding mean values.

\section{Conclusions}

The perspective suggested in this work is based on assumption that the
thermodynamic influence of the environment is omnipresent in the real world;
insulating experimental systems only weakens this influence but cannot exclude
it completely. Although decays have stronger coupling with thermodynamic
environment than quantum steady states, the direct influence of the
environment on invariant properties may be difficult to notice in systems that
are CP-invariant --- the overwhelming majority of quantum systems belong to
this group. In the case of intrinsic CP violations, interference of the
environment is likely to produce the impression of a CPT violation: the decay
of neutral kaons seems to support this view. This violation is only apparent
and does not contradict strict interpretations of the CPT symmetry, which is
fundamental in particle physics.

\section*{Acknowledgments}

The author's work is supported by the Australian Research Council.

\appendix

\section*{APPENDIX: The random phase notation}

This appendix explains the system of notation that is based on random phases
and used in this paper. These notations can be useful in distinguishing pure
and mixed states of quantum mechanics, especially in cases where dealing with
wave functions seems to be more convenient than using density matrices. We
stress that the random phases serve only to provide notional convenience and
do not represent a physical explanation or a theory. Let
\begin{equation}
\left[  \beta\right]  =\exp(-i\theta_{\beta}),\;\;\;\left[  \beta\right]
^{\ast}=\exp(+i\theta_{\beta}),
\end{equation}
where $\theta_{\beta}$ is a random angle uniformly distributed between 0 and
2$\pi.$ When a wave function is multiplied by a random phase, evaluation of
the density matrix involves averaging over the random phases, which is
obviously compliant with the following rule
\begin{equation}
\left\langle \left[  \alpha\right]  ^{\ast}\left[  \beta\right]  \right\rangle
=\left\langle \exp\left(  -i(\theta_{\beta}-\theta_{\alpha})\right)
\right\rangle =I_{\alpha\beta} \label{RPH1}%
\end{equation}
where $I_{\alpha\beta}$ is the identity matrix:
\begin{equation}
I_{\alpha\beta}=\left\{
\begin{array}
[c]{cc}%
1 & \alpha=\beta\\
0 & \alpha\neq\beta
\end{array}
\right.
\end{equation}
Note that the product of two random phases $\left[  \beta_{3}\right]  =\left[
\beta_{1}\right]  \left[  \beta_{2}\right]  $ is also a random phase that is
different from both $\left[  \beta_{1}\right]  $ and $\left[  \beta
_{2}\right]  $ since $\left\langle \left[  \beta_{3}\right]  ^{\ast}\left[
\beta_{1}\right]  \right\rangle =0$ and $\left\langle \left[  \beta
_{3}\right]  ^{\ast}\left[  \beta_{2}\right]  \right\rangle =0.$
Multiplication of a random phase by a fixed phase is still a random phase,
although not an independent one: $\left[  \beta^{\circ}\right]  =e^{-i\phi
}\left[  \beta\right]  $ and $\left\langle \left[  \beta^{\circ}\right]
^{\ast}\left[  \beta\right]  \right\rangle =e^{i\phi}$.

Consider the following examples. The first expression
\begin{equation}
\left|  \psi_{1}\right\rangle =\sum_{\beta}c_{\beta}\left|  \beta
\right\rangle
\end{equation}
represents a pure state, which is a superposition of pure states $\left|
\beta\right\rangle .$ The second expression
\begin{equation}
\left|  \psi_{2}\right\}  =\sum_{\beta}c_{\beta}\left[  \beta\right]  \left|
\beta\right\rangle
\end{equation}
represents a mixed state, which is a mixture of the pure states $\left|
\beta\right\rangle $. Let us evaluate the corresponding density operators to
see the difference
\begin{equation}
\bbrho_{1}=\left|  \psi_{1}\right\rangle \left\langle \psi_{1}\right|
=\sum_{\alpha,\beta}c_{\alpha}^{\ast}c_{\beta}\left|  \beta\right\rangle
\left\langle \alpha\right|
\end{equation}%
\begin{equation}
\bbrho_{2}=\left|  \psi_{2}\right\}  \left\{  \psi_{2}\right|  =\sum
_{\alpha,\beta}c_{\alpha}^{\ast}c_{\beta}\underset{=I_{\alpha\beta}%
}{\underbrace{\left\langle \left[  \alpha\right]  ^{\ast}\left[  \beta\right]
\right\rangle }}\left|  \beta\right\rangle \left\langle \alpha\right|
=\sum_{\beta}c_{\beta}c_{\beta}^{\ast}\left|  \beta\right\rangle \left\langle
\beta\right|  \label{den}%
\end{equation}
where rule (\ref{RPH1}) is applied. The second density operator corresponds to
a mixed state.

In a mixed state, superposition and mixing can be performed over different
indices. For example, the expression
\begin{equation}
\left|  \psi\right\}  =\sum_{s,\beta}c_{s\beta}\left[  \beta\right]  \left|
s\right\rangle
\end{equation}
represents the pure states
\begin{equation}
\left|  \psi_{\beta}\right\rangle =\sum_{s}\frac{c_{s\beta}}{p_{\beta}^{1/2}%
}\left|  s\right\rangle ,\;\;\beta=1,2,...
\end{equation}
that are mixed with the probabilities $p_{\beta}=\Sigma_{s}c_{s\beta}^{\ast
}c_{s\beta}.$

Generally, the random phase notation is equivalent to notations using the
density matrices, although in some cases the random phases can distinguish
different mixed states that have identical density matrices. A good example,
which has been mentioned in many publications \cite{PenroseBook}, can be
expressed in terms of the random phases by the following expressions
\begin{align}
\left|  \psi_{+1}\right\rangle  &  =\frac{\left|  \uparrow\right\rangle
+\left|  \downarrow\right\rangle }{\sqrt{2}},\;\;\left|  \psi_{-1}%
\right\rangle =\frac{\left|  \uparrow\right\rangle -\left|  \downarrow
\right\rangle }{\sqrt{2}}\nonumber\\
\left|  \psi_{2}\right\}   &  =\frac{\left[  \alpha\right]  \left|
\uparrow\right\rangle +\left[  \beta\right]  \left|  \downarrow\right\rangle
}{\sqrt{2}},\;\;\left|  \psi_{3}\right\}  =\frac{\left[  \alpha\right]
\left|  \psi_{+1}\right\rangle +\left[  \beta\right]  \left|  \psi
_{-1}\right\rangle }{\sqrt{2}}%
\end{align}
The first two expressions specify $\left|  \psi_{+1}\right\rangle $ and
$\left|  \psi_{-1}\right\rangle $ as pure states that are different
superpositions of two other pure states $\left|  \uparrow\right\rangle $ and
$\left|  \downarrow\right\rangle $ representing spin. The last two expressions
indicate that $\left|  \psi_{2}\right\}  $ is a mixture of two pure states
$\left|  \uparrow\right\rangle $ and $\left|  \downarrow\right\rangle $ with
equal probability and that $\left|  \psi_{3}\right\}  $ is a mixture of two
other pure states $\left|  \psi_{+1}\right\rangle $ and $\left|  \psi
_{-1}\right\rangle $. \ Note that $\left|  \psi_{3}\right\}  $ is different
from $\left|  \psi_{2}\right\}  $ and this difference is reflected by the
random phase notation. The density operators corresponding to these cases are
evaluated according to (\ref{den}) and, in matrix form, are given by
\begin{align}
\bbrho_{+1}  &  =\frac{1}{2}\left[
\begin{array}
[c]{cc}%
1 & 1\\
1 & 1
\end{array}
\right]  ,\;\;\bbrho_{-1}=\frac{1}{2}\left[
\begin{array}
[c]{cc}%
1 & -1\\
-1 & 1
\end{array}
\right] \nonumber\\
\bbrho_{2}  &  =\frac{1}{2}\left[
\begin{array}
[c]{cc}%
1 & 0\\
0 & 1
\end{array}
\right]  ,\;\;\;\bbrho_{3}\;=\;\frac{1}{2}\left[
\begin{array}
[c]{cc}%
1 & 0\\
0 & 1
\end{array}
\right]
\end{align}

Another interesting example is given by the mixture
\begin{equation}
\left|  \psi\right\}  =p_{\alpha}^{1/2}\left[  \alpha\right]  \left|
1\right\rangle +p_{\beta}^{1/2}\left[  \beta\right]  \left(  \;c_{1}\left|
1\right\rangle +c_{2}\left|  2\right\rangle \;\right)
\end{equation}
of two pure states $\left|  \psi_{\alpha}\right\rangle =\left|  1\right\rangle
$ and $\left|  \psi_{\beta}\right\rangle =c_{1}\left|  1\right\rangle
+c_{2}\left|  2\right\rangle $, where $c_{1}^{\ast}c_{1}+c_{2}^{\ast}c_{2}=1$
and $p_{\alpha}+p_{\beta}=1.$ Note that arithmetically the same expression
\begin{equation}
\left|  \psi\right\}  =\left(  p_{\alpha}^{1/2}\left[  \alpha\right]
+p_{\beta}^{1/2}c_{1}\left[  \beta\right]  \right)  \left|  1\right\rangle
+p_{\beta}^{1/2}c_{2}\left[  \beta\right]  \left|  2\right\rangle
\end{equation}
would be more difficult to interpret. Hence, the random phases need to be
factored out and the summation of different random phases, such as $p_{\alpha
}^{1/2}\left[  \alpha\right]  +p_{\beta}^{1/2}\left[  \beta\right]  ,$ is
generally not permitted. If $c_{2}\rightarrow0$ then $\left|  \psi\right\}  $
becomes
\begin{equation}
\left|  \psi\right\}  =\left(  \;p_{\alpha}^{1/2}\left[  \alpha\right]
+p_{\beta}^{1/2}\left[  \beta\right]  \;\right)  \left|  1\right\rangle
=\left|  \psi\right\rangle
\end{equation}
which, as the mixture of the same pure states $\left|  \psi_{\alpha
}\right\rangle =\left|  \psi_{\beta}\right\rangle =\left|  1\right\rangle ,$
is conventionally interpreted as a pure state $\left|  \psi\right\rangle $. If
we follow this convention, then in this case the sum of the random phases can
be formally interpreted as another random phase $\left[  \gamma\right]  ,$
that is $\left[  \gamma\right]  =p_{\alpha}^{1/2}\left[  \alpha\right]
+p_{\beta}^{1/2}\left[  \beta\right]  $.

In these work we distinguish pure and mixed states by different bra and ket
symbols. Since, as canonical typicality indicates, the difference between pure
and mixed states tends to be blurred for very large quantum systems, using the
same bras and kets for both $\left|  \psi\right\rangle $ and $\left|
\psi\right\}  $ may be more convenient. The random phases notation is a
convenient notation for distinguishing pure and mixed states of quantum
mechanics. In conventional quantum mechanics, the random phase multipliers
$\left[  \alpha\right]  ,$ $\left[  \beta\right]  $ are orthonormal and do not
evolve in time. Spontaneous decoherence, however, corresponds to $\left[
\alpha\right]  $ being the same as $\left[  \beta\right]  $ initially but then
evolving into stochastically independent quantities.


\end{document}